\documentclass[aps,twocolumn,preprintnumbers,showpacs]{revtex4}

\usepackage{color}
\usepackage{epsfig}
\usepackage{amsmath,amssymb}
\usepackage{graphicx}
\usepackage{revsymb}

\newcommand{\lp}{\ell_{\mathrm P}}
\newcommand{\be}{\begin{equation}}
\newcommand{\ee}{\end{equation}}
\newcommand{\bq}{\begin{eqnarray}}
\newcommand{\eq}{\end{eqnarray}}

\newcommand{\mpl}{M_{\rm P}}

\def\H{{\cal H}}

\def\f{\frac}
\def\t{\tilde}

\begin{document}

\title{Loop cosmological implications of a non-minimally coupled scalar field.}
\author{Martin Bojowald\footnote{e-mail address: {\tt bojowald@gravity.psu.edu}}
and Mikhail Kagan\footnote{e-mail address: {\tt mak411@psu.edu}}}
\vspace{.5cm} \affiliation{Institute for Gravitational Physics and Geometry,\\
The Pennsylvania State University,\\ 104 Davey Lab, University
Park, PA 16802, USA} 

\begin{abstract}
Non-minimal actions with matter represented by a scalar field
coupled to gravity are considered in the context of a homogeneous
and isotropic universe. The coupling is of the form $-\f{1}{2}\xi
\phi^2 R$. The possibility of successful inflation is investigated
taking into account features of loop cosmology. For that end a
conformal transformation is performed. That brings the theory into
the standard minimally coupled form (Einstein frame) with some
effective field and its potential. Both analytical and numerical
estimates show that a negative coupling constant is preferable for
successful inflation. Moreover, provided fixed initial conditions,
larger $|\xi|$ leads to a greater number of {\em e}-folds. The
latter is obtained for a reasonable range of initial conditions
and the coupling parameter and indicates a possibility for
successful inflation.
\end{abstract}
\maketitle

\section {Introduction}
Modern astronomical observations gave rise to a number of
important conceptual issues, which can be successfully resolved by
the mechanism called {\em inflation}. According to the
inflationary paradigm \cite{Guth,L&L}, there has to have been an
epoch, when the Universe was undergoing an accelerated expansion;
such that the size of the Universe increased by a factor of about
$e^{60}$ (60 e-foldings) to agree with observations.

This scenario can be realized within the framework of the
Friedmann-Robertson-Walker model, if the matter is represented by
a self-interacting scalar field, the {\em inflaton}, which is
minimally coupled to gravity. The mechanism considers so called
{\em slow-roll approximation}, when the inflaton rolls down its
potential hill towards the potential minimum, but, at the same
time, remains far away from the minimum. It is in the vicinity of
the minimum of the potential, where inflation stops. For the
simplest choice of a quadratic field potential with an arbitrary
mass of the inflaton, one would obtain 60 e-foldings if the
slow-roll regime starts off at $\phi \approx 3 M_{\rm P}$ and ends
when the field is essentially zero. The question as to how the
inflaton gets to that value before the slow-roll phase is left for
a quantum theory to answer.

Without a quantum theory taking into account gravity at hand, one
could heuristically argue along the lines of Linde's chaotic
inflationary scenario \cite{Linde}. The idea was that vacuum
fluctuations of a scalar field would be distributed randomly up to
Planckian energy densities. Regions with sufficient values of
$\phi$ would inflate resulting in a flat FRW-universe. We shall
see that, however insightful, this paradigm imposes severe
restrictions on the likelihood of inflation in non-minimally
coupled models, with a coupling of the form $-\f{1}{2}\xi \phi^2
R$ \cite{Conf_NMC,Conf_NMC1}. Specifically, the coupling constant
needs to be very small for successful inflation: $\xi < 10^{-12}$
for positive coupling and $|\xi|<10^{-3}$ for negative coupling.

Loop Quantum Cosmology (LQC) \cite{LivRev} has over the last years
resolved some long-standing issues of cosmology and general
relativity in whole and suggested an alternative mechanism to
chaotic inflation. As has been shown in \cite{ModFried,2ndCorr},
LQC leads to effective quantum modifications of the classical
Friedmann equations, such that gravity becomes repulsive at small
scales and introduces an `anti-friction' term into the
Klein-Gordon equation for the scalar, thus driving the inflaton up
its potential hill \cite{closedinflation,ChaInf}. Therefore even
small vacuum fluctuations of $\phi$ will be amplified during the
quantum-corrected (effective) phase of evolution. In the light of
this, much smaller field initial values are required to start a
successful classical inflation. This fact significantly relaxes
the restrictions on the coupling constant of a non-minimally
coupled theory. Moreover, it is possible to study the inflaton's
`climbing-up' systematically, provided some initial fluctuations
occur in the field and its canonically conjugate momentum.

The paper is organized as follows. We first review the
conventional model for a non-minimally coupled scalar field. Then,
following \cite{Conf_NMC,Conf_NMC1}, we perform a conformal
transformation, which allows us to recast the action in the
minimally coupled form in terms of a redefined field variable with
some effective potential. In section three, we investigate the
slow-roll regime and derive the number of e-folds as a function of
the value of the inflaton right before inflation. Then, in section
four, we turn to investigating the inflaton's `climbing', that is
how small initial fluctuations get amplified during the
quantum-corrected phase and the following classical evolution. The
estimates for the maximum field value are mostly obtained
analytically, up to a certain point when one is forced to solve a
transcendental equation. The results are then compared with the
numerical solutions of the Hamilton equations of motion in section
five. The comparison indicates that the analytical method provides
a decent precision and hence can be efficiently used for the
analysis of the range of initial conditions and parameters that
lead to a sufficiently long inflationary phase.

\section{Action for a non-minimally coupled scalar and conformal transformation}
The simplest form of a non-minimally coupled action is given by
\be \label{Action} S[g_{ab}, \phi]=\int{d^4 x \sqrt{-g} \left(
\frac{f(\phi)}{2 \kappa} R - \frac{1}{2}g^{ab}\nabla_a \phi
\nabla_b \phi - U(\phi) \right)}
\ee
where $g \equiv \det(g_{ab})$, $\kappa = 8 \pi G$, $R$ is the
scalar curvature of the metric and $U(\phi)$ is the
self-interacting potential of the matter scalar field $\phi$. The
curvature term is coupled to matter through the coupling function
$f(\phi)$ and is manifestly Lorentz invariant. There are some
obvious restrictions on the coupling function. First of all, as we
shall see, it enters the symplectic structure for the
gravitational (connection) variables. Thus it must not vanish
anywhere. Moreover, one can view this function as defining an
effective gravitational constant $\kappa$, in the sense
\be \label{kappaEff} \kappa_{\rm eff} = \kappa/f(\phi) \ee
Hence we require $f(\phi)$ to be always positive. Note also that
the scalar field is expected to change sign, e.g. during the
reheating epoch. It is then natural to restrict $f(\phi)$ to be an
even function of $\phi$. Finally, the coupling function must
satisfy the limit $f(\phi) \rightarrow 1$ when $\phi \rightarrow
0$, so that the standard normalization is recovered in the absence
of a scalar field.

Given the restrictions above, the simplest form of the coupling
function is
\be \label{CFcn} f(\phi)=1-\sigma \phi^2.\ee
(One can think of this expression as the Taylor expansion up to
the quadratic order. We are interested in the case of a weak
coupling, so the latter should be a good approximation. Note also
that coupling of higher degree would require a dimensionful
coupling constant, while $\xi$ is unitless.) In fact, this is the
form of coupling function that is normally considered in the
literature (see, for example, \cite{Carroll}). We will refer to
$\sigma$ as the {\em coupling strength} (not to be confused with
the {\em coupling constant} $\xi\equiv \f{\sigma}{\kappa}$ of
\cite{Conf_NMC}). Note that zero $\sigma$ corresponds to the case
of minimal coupling. If $\sigma>0$, the coupling function can, in
principle, become zero and even negative if the value of the
scalar field exceeds the critical value
\be \label{phi_crit}\phi_{\rm crit}:=\f{1}{\sqrt{|\sigma|}} \ee
We have deliberately written the absolute value of the coupling
strength because, as we shall see later on, it is also of
importance in the case of negative coupling.

In this paper, we restrict our attention to a flat homogeneous and
isotropic model, described by the FRW-metric
\be
ds^2=-dt^2+a(t)^2 d\vec{r}^{\,2}
 \ee
Using a conformal transformation
\be \label{CT} \t g_{ab} := f(\phi) g_{ab}\ee
the action (\ref{Action}) can be recast in the form
\cite{Conf_NMC}
\be \label{action}
S[\t g_{ab},\phi]=\int{dt \left[ \f{3}{\kappa}(a^2 \ddot a + a
\dot a^2)+a^3\left(\f{1}{2}F(\phi)^2\dot\phi^2-\t
V(\phi)\right)\right]}
\ee
where we have introduced the effective potential
\be \label{Veff}
\t V(\phi):=\f{U(\phi)}{f(\phi)^2} \ee
and
\be F(\phi)^2:=\f{1-\sigma \phi^2 (1-\f{6
\sigma}{\kappa})}{f(\phi)^2} \approx \f{1}{f(\phi)}. \ee
The last approximation holds as long as we are interested in weak
coupling such that $\sigma \sim (10^{-3}\div 10^{-2}) \kappa$;
then $\left( 1- \f{6 \sigma}{\kappa}\right)$ can be set to 1 in
this order of magnitude. Note that in the Hamiltonian formulation,
there is a similar canonical transformation under which
$F(\phi)^2:=\f{1}{f(\phi)}$ exactly \cite{nonMin}. Consequently,
the relation between the scalar fields (\ref{phiRel}), derived
below, also becomes exact, which considerably simplifies the
analysis even if $\f{6 \sigma}{\kappa} \neq 1$. The last step,
that will bring the kinetic term into its canonical form (Einstein
frame), is to redefine the field
\bq \label{phiDef} \t \phi &:=&\int{d\phi F(\phi)} \approx
\int{\f{d\phi}{\sqrt{1-\sigma
\phi^2}}} \nonumber\\ &=&\f{1}{\sqrt{\sigma}}\left\{%
\begin{array}{ll}
    \sin^{-1}(\sqrt{\sigma\phi}), & \hbox{if}  \quad \sigma > 0\\
    \sinh^{-1}(\sqrt{|\sigma|}\phi), & \hbox{if} \quad \sigma < 0 \\\end{array}%
\right. \eq
Note that for a positive $\sigma$ the quantity $\sqrt{\sigma}\phi$
must be less than unity, for $f(\phi)$ must not vanish and, in
fact, be positive. From (\ref{phiDef}) we get
\be \label{phiRel}
\sqrt{|\sigma|}\phi \approx\left\{%
\begin{array}{ll}
    \sin(\sqrt{\sigma}\t\phi), & \sigma > 0\\
    \sinh(\sqrt{|\sigma|}\t\phi), & \sigma < 0 \\\end{array}%
\right. \ee

Aiming at a loop quantization, we now have all the ingredients to
proceed to the Hamiltonian formulation of the theory. For that end
we introduce the phase space variables:
\be |\t p|:=a^2, \quad \t c:=\gamma \dot a, \quad \t \phi \quad
{\rm and} \quad \t \pi:=p^{3/2}\dot{\t \phi} \ee
here $\t p$ and $\t c$ are the gauge-invariant triad and
connection components respectively, $\gamma$ is the
Barbero-Immirzi parameter \cite{AshVarReell,Immirzi}, and $\t \pi$
is the field momentum. Generally, the sign of $\t p$ determines
the orientation of the triad. From now on we assume that $\t p>0$.
Again, the tilded variables stand for quantities written in the
Einstein frame. For notational convenience, however, we will omit
tildes until transforming back to the Jordan frame. In terms of
newly defined variables the action (\ref{action}) takes the form
\be \label{newAction} S=\int{dt \left[ \frac {3}{\kappa \gamma} p
\dot{c} + \pi \dot{\phi} - \H \right]}.
\ee
with the Hamiltonian
\be \label{Ham}
\H = -\f{3}{\kappa \gamma^2} \sqrt{p} c^2 + \f{\pi^2}{2 p^{3/2}} +
p^{3/2} V(\phi)
\ee
From (\ref{newAction}) we see that the pairs $\{p,c\}$ and
$\{\phi,\pi\}$ are indeed canonically conjugate variables with the
Poisson brackets
\be \label{PB} \{c,p\}=\frac{\kappa \gamma}{3}, \quad
\{\phi,\pi\}=1\ee
Before proceeding to inflation, let us comment on a
phenomenological relation between the Einstein and Jordan frames.
More specifically, eventually we are interested in the number of
e-folds during the inflationary epoch
\be \t N:=\ln\left(\f{\t a_e}{\t a_b}\right)\ee
where the subscripts stand for the initial (before inflation) and
final (after inflation) values of the scale factor $\t a$. Suppose
we have solved the equations of motion in the Einstein frame and
obtained $\t a_b$ and $\t a_e$. Then it is clear from (\ref{CT})
that the scale factors in the two frames are related by
\be \label{aRel}
\t a = \sqrt{f(\phi)} a\ee
Therefore the Jordanian number of e-folds is
\be N \equiv \ln\left(\f{a_e}{a_b}\right) =
\ln\left(\f{\sqrt{f(\phi_b)}\t a_e}{\sqrt{f(\phi_e)}\t
a_b}\right)= \t N +
\f{1}{2}\ln\left(\f{f(\phi_b)}{f(\phi_e)}\right) \ee
At the end of inflation, the scalar field is essentially zero,
while at the beginning of inflation it has its maximum value
$\phi_{\rm max}$. Thus, using (\ref{CFcn}), we can explicitly
write
\be \label{Nrel} N=\t N + \f{1}{2} \ln(1-\sigma \phi_{\rm
max}^2)\ee
In principle, the beginning of inflation may occur when the scalar
field is close to its critical value, such that $|\sigma \phi_{\rm
max}| \approx 1$. In this case, a negative coupling $\sigma<0$
would yield a negligible correction to the righthand side of
(\ref{Nrel}). A positive coupling, on the contrary, might give a
significant contribution resulting in a substantial difference
between the Jordanian and Einsteinian number of e-folds.
\section{Inflation}
In this section, we consider the inflationary mechanism called
{\em slow-roll approximation} \cite{L&L}. According to this model,
the inflaton rolls down its potential hill towards the potential
minimum, but, at the same time, remains far away from the minimum.
In such a regime, a scalar field behaves as a cosmological
constant. It is in the vicinity of the minimum of the
potential, where inflation stops. We compute the number of e-folds
as a function of the initial (maximum) value of the scalar field.
We shall see that the result is generic irrespective of the sign
of the coupling strength $\sigma$.

We start with investigating the equations of motion generated by
the transformed Hamiltonian (\ref{Ham}).
\bq \label{HamEq}
\dot p &=&2 \sqrt p \f {c}{\gamma}  = \sqrt{\f{2 \kappa}{3}}
\sqrt{\f{\pi^2}{p}+2 p^2 V(\phi)} \nonumber\\
\dot \phi &=& \f{\pi}{p^{3/2}} \\
\dot \pi &=& -p^{3/2} V^{\prime} (\phi) \nonumber
\eq
where the prime denotes a $\phi$-derivative. In the first equation
we have eliminated the connection $c$, using the Hamiltonian
constraint (\ref{Ham}) $H \approx 0$. We shall make an extensive
use of these equations later on. For the moment, however, it is
more convenient to consider the conventional Friedmann equations.
The constraint equation, rewritten in terms of the scale factor
and its time derivative, reads
\be \H=-\frac{3}{\kappa}a \dot a^2+\frac{a^3}{2}\dot \phi^2+a^3
V(\phi) \approx 0 \ee
Dividing by $a^3$, one gets the Friedmann equation:
\be \label{Fri} H^2=\frac{\kappa}{3}\left[\frac {\dot
\phi^2}{2}+V(\phi)\right] \ee
Here $H \equiv \dot a/a$ is the Hubble rate. Using the $\dot
\pi$-equation we derive the Klein-Gordon equation
\be \label{KG} \ddot \phi+3 H\dot \phi+V^\prime(\phi)=0 \ee
Finally, the Raychaudhuri equation can be obtained by taking the
time derivative of (\ref{Fri}) and substituting $\ddot \phi$ from
(\ref{KG}):
\be \label{Ray} \dot H=-\frac{\kappa}{2}\dot \phi^2 \ee
From now on we stick with the slow-roll approximation that would
ensure inflation. The potential domination (small kinetic terms)
\be \label{potDom} \dot \phi^2 \ll V(\phi), \quad |\ddot \phi| \ll
|3 H \dot \phi| \ee
is provided, if we require the slow-roll parameters to be small:
\bq
\epsilon&=&\f{1}{2 \kappa} \left( \f{V^\prime}{V}\right)^2 \ll 1
\nonumber \\
\eta&=&\f{1}{\kappa} \left( \f{V^{\prime \prime}}{V}\right) \,\,
\ll 1
 \eq
With the assumptions (\ref{potDom}) the Friedmann and Klein-Gordon
equations can be rewritten as
\bq \label{SR_Frie} H^2 \approx \f{\kappa}{3}V(\phi) \nonumber \\
3 H \dot \phi \approx -V^\prime(\phi) \eq
whereas the Raychaudhuri equation simply implies  $\dot H \approx
0$, i.e. an exponential expansion of the scale factor. Dividing
the first of the equations (\ref{SR_Frie}) by the second one, we
can express the Hubble rate as
\be H\approx -\kappa \dot \phi \f{V}{V^\prime} \ee
The number of e-folds is then given by
\bq \label{Ngen}
N&=&\int_{t_b}^{t_e}{H dt} \approx -\kappa
\int_{t_b}^{t_e}{\f{V}{V^\prime} \dot {\phi} dt} \nonumber \\
&=&-\kappa\int_{\phi_b}^{\phi_e}{\f{V}{V^\prime} d\phi}
\eq
where $\phi_b$ and $\phi_e$ are the values of the inflaton before
and after inflation respectively. The last expression tells us
that as soon as the shape of the potential is specified, one can
compute the number of e-folds. Therefore, it is now pertinent to
analyze the form of the effective potential given by (\ref{Veff}).
The most common choice is either quadratic or quartic potential
\be U(\phi)=\f{m^2}{2}\phi^2, \quad {\rm or} \quad
U(\phi)=\f{\lambda}{4}\phi^4 \ee
The corresponding effective potentials are
\bq V(\phi)&=&\f{m^2}{2 \sigma}
\f{\tan^2(\sqrt{\sigma}\phi)}{\cos^2(\sqrt{\sigma}\phi)}, \quad
 {\rm and} \nonumber \\ V(\phi)&=&\f{\lambda}{4 \sigma^2}
\tan^4(\sqrt{\sigma}\phi) \eq
if $\sigma$ is positive, and
\bq V(\phi)&=&\f{m^2}{2 |\sigma|}
\f{\tanh^2(\sqrt{|\sigma|}\phi)}{\cosh^2(\sqrt{|\sigma|}\phi)}, \quad {\rm and} \nonumber \\
 V(\phi)&=&\f{\lambda}{4 \sigma^2}
\tanh^4(\sqrt{|\sigma|}\phi) \eq
if $\sigma$ is negative.

\begin{figure}
\begin{center}
\includegraphics[width=9cm,height=6.0cm]{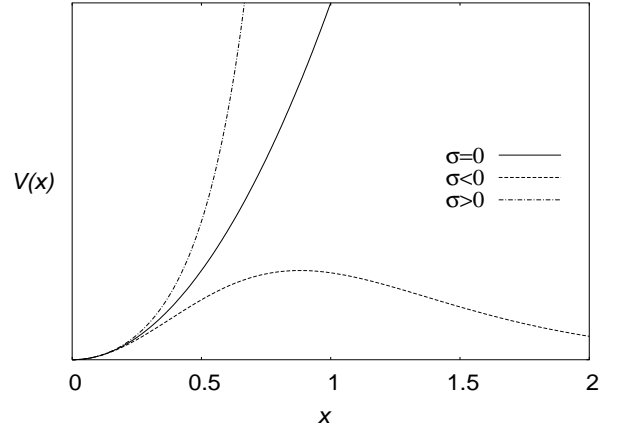}
\end{center} \vskip-0.5cm \caption{The effective potential
for $\sigma<0$ (upper curve) and $\sigma>0$ (lower curve). The
solid curve represents the original quadratic potential. The
argument of the potential is $x\equiv \sqrt {|\sigma|} \phi$ for
$\sigma \neq 0$ and $x=\phi$ for $\sigma=0$.} \label{quadratic}
\end{figure}

Fig. \ref{quadratic} shows the effective potentials for the
quadratic original potential. If $\sigma>0$ (Fig.
\ref{quadratic}), the effective potential is steeper than
quadratic (punched line) and diverges at
$\sqrt{\sigma}\phi=\f{\pi}{2}$ thus keeping the inflaton away from
that point. Note that this is exactly the value of $\phi$ at which
the coupling function goes to zero. On the contrary, the effective
potential for a negative $\sigma$ is bounded from above, goes
below the quadratic potential (punched line) and has a maximum at
$\sqrt{|\sigma|}\phi=\ln(1+\sqrt{2})$. Remarkably, both these
values of the tilded scalar correspond to the critical value of
the original field (\ref{phi_crit}). The existence of the maximum
for a negative coupling constant implies that if the inflaton
exceeds the critical value, it will roll down to the right,
towards infinite values. The latter is unacceptable, as the
inflaton is supposed to dissipate to zero after inflation. We
therefore should restrict our attention to $\phi<\phi_{\rm crit}$
for both positive and negative couplings.

Let us now retrieve the tildes over the conformally transformed
variables, to distinguish them from the original ones, and
calculate the number of e-folds
\bq \label{Ngen}
\t N&=& -\kappa\int_{\t\phi_b}^{\t\phi_e}{\f{\t V}{\t V_{,\t\phi}}
d\t\phi}
\nonumber \\
&=&-\kappa\int_{\phi_b}^{\phi_e}{\f{\t V}{\t V_{,\phi}}
F^2(\phi) d\phi}\\
&=&-\kappa\int_{\phi_b}^{\phi_e}{\f{1}{f(\phi)}\f{d\phi}{\ln\left(
\f{U(\phi)}{f^2(\phi)}\right)^\prime}} \nonumber
\eq
If the original potential is quadratic $U(\phi)=\f{1}{2}m^2
\phi^2$, then (\ref{Ngen}) yields
\be \t N=-\f{\kappa}{2}\int_{\phi_b}^{\phi_e}{\f{\phi
d\phi}{1+\sigma \phi^2}}=\f{\kappa}{4 \sigma} \ln \left|
\f{1+\sigma \phi_b^2}{1+\sigma \phi_e^2}\right| \ee
Neglecting $\phi_e$ as compared to $\phi_b \equiv \phi_{\rm max}$,
and using (\ref{phiRel}), the number of e-folds in the Einstein
frame is
\bq \label{Nquad} \t N & \approx & \f{\kappa}{4 \sigma}\ln \left|
1+\sigma \phi_{\rm max}^2\right| \nonumber \\
&=& \f{\kappa}{4 \sigma} \left\{%
\begin{array}{ll}
    \ln |1+\sin^2(\sqrt{\sigma}\t\phi_{\rm max})|, \quad &\sigma > 0\\
    \ln |1-\sinh^2(\sqrt{|\sigma|}\t\phi_{\rm max})|, \quad &\sigma < 0\\
\end{array}%
\right. \eq
Note that in the limit $\sigma \rightarrow 0$, the number of
e-folds reduces to that of the minimally coupled scalar field.
Furthermore, considering the Taylor expansion of (\ref{Nquad})
\be \f{1}{\sigma}\ln{|1+\sigma x^2|} = x^2 - \f{1}{2}\sigma x^4
+... , \nonumber
\ee
and the inequality $|\sinh(x)|\geq |\sin(x)|$, we conclude that,
provided the same initial field, negative $\sigma$ is preferable
for successful inflation.

Following a similar derivation for a quartic original potential
$U(\phi)=\f{\lambda}{4}\phi^4$, we obtain the number of e-folds
\be \t N=\f{\kappa}{8 |\sigma|}
\left\{%
\begin{array}{ll}
       \sin^2(\sqrt{\sigma}\t\phi_{\rm max}), \quad & \sigma>0 \\
       \sinh^2(\sqrt{|\sigma|}\t\phi_{\rm max}), \quad & \sigma<0
\end{array}%
\right. \ee
As before, in the limit $\sigma \rightarrow 0$ the expression
agrees with that of the minimally coupled model. Furthermore, a
negative coupling constant again provides a greater number of
e-folds.

As was mentioned before, the question as to how the inflaton gets
at $\phi_{\rm max}$ is not addressed in the classical theory. In
the next section, we shall estimate $\phi_{\rm max}$ on the
grounds of LQC.
\section{Estimation of maximum inflaton value}
In the framework of LQC, the self interacting scalar field,
inflaton, arises as a microscopic vacuum fluctuation that is
driven up its potential well by quantum effects. In the isotropic
and homogeneous context, it is indeed quantum corrected
(effective) Friedmann equations \cite{ModFried,2ndCorr} that have
been shown to govern the evolution of the Universe in the
semiclassical regime, when the scale factor $a_0<a<a_*$. The
boundedness of the geometrical density operator \cite{geomDens}
manifests itself in the form of ``repulsive'' effects of gravity
at Planckian scales \cite{LivRev} and leads to an anti-friction
term in the effective Klein-Gordon equation
\cite{closedinflation}.

Practically, the process of the inflaton's `climbing-up' can be
broken into two stages: i) effective, driven by quantum
modifications and consecutive ii) classical phase, conditioned by
the field momentum gained in the effective regime. Both stages
will be assumed to be {\em kinetic-dominated}, that is they occur
relatively rapidly, and the potential terms, whenever they appear
along with kinetic ones, are dominated by the latter. In other
words, $\phi$ almost reaches its maximum before the potential
effects become appreciable.

\subsection{Effective phase}
This phase starts with some initial conditions and ends when the
scale factor $a\approx a_*$, a characteristic scale introduced in
Eq.(\ref{qdef}) below. We should say that the idea behind the
estimate for $\t N$ used in this paper is analogous to that in
\cite{FieldEst} except that the semiclassical phase ends {\em
before} the inflaton reaches its maximum. Nor do we assume any
specific asymptotic limit of the geometrical density spectrum.

 The recipe provided by LQC as to how to proceed
directly to an effective description is the following. In the
Hamiltonian constraint, one is to replace all negative powers of
triad $p$ with appropriate factors of the spectrum of the inverse
volume operator \cite{dQ}
\be \label{HamSC} \f{1}{p^{3/2}} \rightarrow d_j  = \f{D\left( q
\right)} {p^{3/2}},\ee
where%
\bq \label{defD}
D(q) &=& q^{3/2} \left\{  \frac {8}{11} \right. \left[
(q+1)^{11/4}-|q-1|^{11/4}\right ] \nonumber \\ &-& \frac{8}{7} q
\left. \left[ (q+1)^{7/4}-{\rm sgn}(q-1) |q-1|^{7/4}\right ]
\right \}^6 \eq
and%
\be \label{qdef}%
q \equiv \f{p}{p_*} \quad {\rm with}\quad p_* \equiv a_*^2:=\f{8
\pi \gamma j \mu_0}{3} \lp^2 \ee
The Planck length is defined as $\lp \equiv \sqrt{\kappa/8 \pi}
\equiv \mpl^{-1}$. The asymptotic behavior of the density operator
follows from the equation above. Not only does $d_j$ remain
bounded for $p \rightarrow 0$, it becomes proportional to a
positive power of the scale factor. Specifically, $D(p/p_*)
\propto p^{15/2}$ and $d \propto p^{6}$, thus making the matter
Hamiltonian (\ref{defD}) well behaved. On the other hand, as $p$
goes to infinity, one  recovers the classical dependence:
$D(p/p_*) \approx 1$ whereas $d \approx p^{-3/2}$.

Note, however, that this Hamiltonian constraint does not account
for higher order perturbative corrections that proved to be
important for small values of the parameter $j$
\cite{2ndCorr,SemiSV,Date}. More precisely, one has to consider
the modified Hamiltonian
\be \label{HO_Ham} \H_{\rm eff}=-\frac{3}{\kappa \gamma^2
\mu_0^2}\sqrt{p}\,{\rm sin}^2(\mu_0 c)+\frac{1}{2}d_j(a)
\pi^2+V(\phi)
\ee
When $\mu_0 c \ll \pi/2$ one recovers the Hamiltonian (\ref{Ham}).
The discreteness corrections become important, when the matter
part of the Hamiltonian is of order of the critical density
\be \label{critDens} \rho_{\rm cr}=3/\kappa \mu_0^2\gamma^2 a^2
\ee
corresponding to the maximum value of the gravitational part,
attained when ${\rm sin}(\mu_0 c)=1$.

Let us focus on the case, when the initial conditions are
such that $\mu_0 c \ll \pi/2$. We shall investigate under what
restrictions this inequality will hold during the evolution within
the effective regime. The effective counterparts of equations of
motion (\ref{HamEq}) take the form
\bq \label{HamEqSc}
\dot p &=&  \sqrt{\f{2 \kappa}{3}} \sqrt{D\left(
q\right)\f{\pi^2}{p}+2 p^2 V(\phi)} \nonumber\\
\dot \phi &=& D\left(
q \right) \f{\pi}{p^{3/2}} \\
\dot \pi &=& -p^{3/2} V^{\prime} (\phi) \nonumber
\eq
As usual we have eliminated the connection variable using the
constraint equation
\be \label{ceq} \frac{c}{\gamma}=\sqrt{\frac{\kappa}{6}}
\sqrt{\frac{\pi^2}{p^2}D(q) +2 p V(\phi)} \ee
Note that within the effective regime, the field is very small
($\sigma \phi^2 \ll 1$) and the non-triviality of the coupling can
be neglected. Let us therefore set $f(\phi)$ to unity.

From now on we take the quadratic potential \be
V(\phi)=\frac{1}{2} m^2 \phi^2, \ee $m$ being the field mass.
Consider the initial stage of the effective evolution, when the
scale factor ranges from $p \approx p_0 \equiv \frac{4 \pi \gamma
\mu_0}{3}$ to $p=p_*$. For a small value of the inflaton ($\phi$
being of order 0.1), we see that in (\ref{ceq}) the kinetic term
overwhelmingly dominates over the potential term, provided the
field mass $m=10^{-6} \mpl$. Thus for the sake of estimate of the
maximum of $\mu_0 c$ we can discard the last term. Moreover, we
notice that multipliers in the first term can be split into
``fast'' and ``slow'' ones. Formally, $\pi$ is a slow variable, as
the righthand side of $\dot \pi$-equation in (\ref{HamEqSc}) is
small due to a factor of $m^2$ in the potential term, whereas $p$
is a fast variable. As far as $D(q)$ is a large positive power of
$p$, the $D(q)$-factor in (\ref{ceq}) is the fastest. More
rigorously, this statement can be cast as
\be \left|\frac{\dot D(q)}{D(q)}\right| \gg \left|\frac{\dot
p}{p}\right| \gg \left|\frac{\dot \pi}{\pi} \right|. \ee
Although, $\phi$ may become as fast as $p$, especially in the end
of the effective phase, it is not used in the following
derivation. The time derivative of $c$ can be approximately
written as
\be  \frac{\dot c}{\gamma} \approx \sqrt{\frac{\kappa}{6}}
\frac{d}{dt} \left( \frac{\pi}{p} \sqrt{D(q)} \right) \approx
\sqrt{\frac{\kappa}{6}} \frac{\pi}{p} \frac {d(\sqrt{D(q)})}{dt}
\ee
This expression vanishes when \be \dot D = \frac{d D(q)}{d p} \dot
p=0,\nonumber \ee which is equivalent to
\be \frac{d D(q)}{d q} = 0 \nonumber \ee
The solution to the last equation is $q_{\rm max} \approx 0.97$,
which yields $\sqrt{D(q_{\rm max})}/q_{\rm max} \approx 1.2$.
Neglecting the change in $\pi$ and taking its initial value, we
obtain the estimate for the maximum value of $\mu_0 c$
\be \mu_0 c_{max} =
\sqrt{\frac{\kappa}{6}}\f{\sqrt{D(q_{max})}}{q_{max}}\mu_0 \gamma
\frac {\pi_i}{p_*} = \f{0.3}{j} \times \pi_i \ee
This remarkable result tells us that, provided the aforementioned
restrictions, the maximum value of $\mu_0 c$, and hence the
maximum matter density, depends only on the initial field
momentum, regardless the initial scale factor and field itself.
Specifically $\rho_{\rm max}$ is linearly proportional to
$\pi_i^2$. Thus the approximation $\mu_0 c \ll \pi/2$ remains
valid throughout the effective regime, if
\be \label{InitPi} \pi_i \ll 5j \lp \ee
Note that a similar derivation goes through, if the modified
Hamiltonian (\ref{HO_Ham}) is used. At a given scale factor, the
matter density must not exceed the critical value
(\ref{critDens}). The condition on the initial momentum
(\ref{InitPi}) then becomes
\be%
\pi_i \leq \sqrt{\f{2 \kappa}{3 D(q_{\rm max})}} \approx 3.45 j
\lp \nonumber \ee
This condition is not very restrictive and compatible with
characteristic values of initial field momentum fluctuations
derived from the Heisenberg uncertainty principle, provided
initial field fluctuations of order $10^{-1} \mpl$. For a more
detailed analysis see \cite{initFluc}.

Now suppose that the effective evolution starts at some initial
data $p=p_i, \phi=\phi_i, \pi=\pi_i$. If $\pi_i$ satisfies the
condition (\ref{InitPi}), we can safely use the approximate
Hamiltonian (\ref{HamSC}) and Hamilton equations (\ref{HamEqSc}).
In order to estimate the maximum value of the inflaton at the end
of the effective phase, it is convenient to rewrite the equations
above in terms of the following dimensionless variables
\be \label{units} (m t) \rightarrow t, \quad x:=\frac{p}{p_*},
\quad y:=\frac{\phi}{\mpl}, \quad z:=\frac{\pi}{m \mpl p_*^{3/2}}
\ee
In these variables the equations of motion (\ref{HamEqSc}) take
the form
\bq
 \label{xdot} \dot x&=& \frac{16 \pi}{3}\sqrt{D(x)\frac{z^2}{x}+x^2 y^2} \\
 \label{ydot} \dot y&=&D(x) \frac{z}{x^{3/2}} \\
 \label{zdot} \dot z&=&-x^{3/2} y
\eq
We are going to get the approximate values for $y$ and $z$ (hence
$\phi$ and $\pi$) at the end of the effective phase. In the same
fashion, as was done before, we can integrate Eq. (\ref{ydot}),
treating $z$ as a slow variable and fixing it at a constant value,
that corresponds to $z_i$. Dividing (\ref{ydot}) by (\ref{xdot}),
we express the derivative
\be%
\f{dy}{dx}\approx \f{D(x)}{x} \ee
then the integration yields
\be \label{phi_eff} y(x) \approx y_i+\sqrt{\frac{3}{2 \kappa}}
\int_{x_i}^{x}{\frac{\sqrt{D(x^\prime)}}{x^\prime}dx^\prime}.
\ee Similarly
\be \frac{d(z^2)}{dx}  \approx - \sqrt{\frac{3}{4 \pi}} \frac{x^2
y(x)}{\sqrt{D(x)}} \ee
Where we again have neglected the potential term in (\ref{xdot}).
Integrating we get
 \be z^2(x) \approx z_i^2-\sqrt{\frac{3}{4 \pi}}
 \int_{x_i}^{x} \!\!\!\! {\frac{x^{\prime^2}dx^\prime}{\sqrt{D(x^\prime)}}
\left( y_i+\sqrt{\frac{3}{16 \pi}} \int_{x_i}^{x^\prime} \!\!\!\!
{\frac{\sqrt{D(x^{\prime \prime})}}{x^{\prime \prime} }dx^{\prime
\prime} }\right)}
\ee
Direct calculation shows that given reasonable initial conditions,
the changes in the field and its momentum are indeed small and can
be neglected when computing the value of the maximum matter
density. In the dimensionful variables, the value of the inflaton
at the end of the effective regime, $\phi_0$, is
\be \label{phinode} \phi_0 \approx \phi_i+ \sqrt{\f{3}{2 \kappa}}
\int_{q_i}^{q_0}{\f{\sqrt{D(q)}}{q}dq}\ee
Assuming $q_i=\f{1}{2 j}=0.1$ and $q_0\approx 1$, the change in
the scalar field is
\be \Delta \phi_{\rm eff}\equiv \phi_0-\phi_i = 0.14 \mpl \ee
The subscript `eff' stands for effective. Remarkably the growth of
the inflaton does not depend on its initial momentum at all!

We should now take the obtained values for $p_0$ and $\phi_0$, as
well as $\pi_0\approx \pi_i$, to be the initial data for the
classical phase and estimate how high the inflaton will get.

\subsection{Classical phase}
When the scale factor significantly exceeds $a_*$, the quantum
corrections become negligible and the evolution is very well
described by the classical equations (\ref{HamEq}). Note, however,
that we can no longer consider the inflaton to be small and should
use the effective potential.

\begin{figure}
\begin{center}
\includegraphics[width=7.5cm,height=5.0cm]{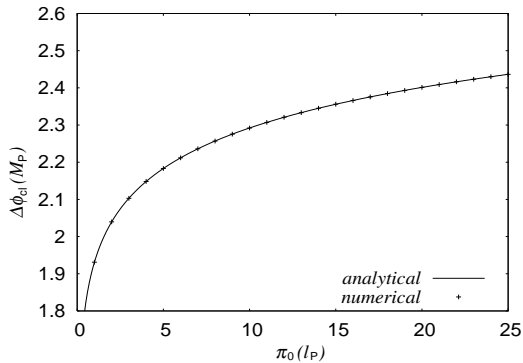}
\end{center}
\vskip-0.5cm \caption{Maximum change of the inflaton as a function
of the initial field momentum for $\sigma=0$ and $p_0=p_*$. The
analytical curve is the Lambert function defined in
(\ref{apprTrans}), whereas the numerical points are obtained as
solutions to the equations of motion (\ref{HamEqSc}).
}\label{fig2}
\end{figure}

The end of the `climbing' phase, i.e. the beginning of inflation,
occurs at $\dot \phi \propto \pi =0$, which does not correspond to
any a-priori fixed scale factor. This is exactly the reason why
one cannot use the technique of the previous sub-section. Instead
one can again assume kinetic domination almost up to the top
\cite{Kevin}, where $\phi=\phi_{\rm max}$. In other words, the
inflaton changes insignificantly after the kinetic terms become
less than the potential ones near $\phi_{\rm max}$. Note that we
neglect the potential only when it appears along with a kinetic
term; we should still keep the right-hand side of the
$\dot\pi$-equation. In the light of this, we eliminate the time
derivatives in Eqs. (\ref{HamEq}):
\bq \label{HamEqCl}
\f{d p}{d\phi} &=&  \sqrt{\f{2 \kappa}{3}} p \nonumber\\
\f{d \pi}{d \phi} &=& -\f{p^3}{\pi}V^\prime(\phi) %
\eq
Solving the first equation for the scale factor as a function of
$\phi$
\be p(\phi)= p_0 \exp\left\{ \sqrt\f{2 \kappa}{3} (\phi-\phi_0)
\right\}\ee
and substituting this expression into the second equation, we
obtain
\be \f{d \pi}{d \phi}=-\f{p_0^3}{\pi}\exp{\left\{\sqrt{6
\kappa}(\phi-\phi_0)\right\}}V^\prime(\phi) \ee
Finally, separating the variables in the last equation and
integrating it from $\pi~\!\!\!=\!\pi_0$ (beginning of the
classical climbing) to $\pi~\!\!=\!0$ (end of the classical
climbing), we arrive at a transcendental equation for $\phi_{\rm
max}$:
\be \label{trans}
\f{\pi_0^2}{2 p_0^3}=\int_{\phi_0}^{\phi_{\rm max}}
{\exp{\left\{\sqrt{6
\kappa}(\phi-\phi_0)\right\}}V^\prime(\phi)}d\phi \ee
Before giving numerical solutions, let us draw some analytical
conclusions from this equation. As the effective potential is
steeper for a positive $\sigma$, we should expect a smaller
maximum value of the inflaton in this case than for a negative
coupling. Note also that in the integrand of (\ref{trans}) the
exponential is the fastest growing function, in the same sense as
before. Therefore, when integrating, the potential factor can be
taken out of the integral and evaluated, for the sake of estimate,
at the upper limit. We now restrict our attention to the case of
quadratic original potential and consider the Taylor expansion of
the effective potential
\be V(\phi)=\f{1}{2}m^2\phi^2 (1+\f{5}{3}\sigma \phi^2+
O(\phi^4))\label{linearSigma}\ee
Then the transcendental equation (\ref{trans}) can be
approximately rewritten as
\be \label{apprSigma}%
 \f{9}{5 y^3}(A {\rm e}^{-y}-y)=\f{\sigma}{\kappa} \ee
where we have defined
\be%
y:=\sqrt{6 \kappa}(\phi_{\rm max}-\phi_0), \quad A:=\f{3 \kappa
\pi_0^2}{m^2 p_0^3} \ee
Let us first set $\sigma$ to zero and denote the corresponding
solution $\bar y$, i.e. $\bar y$ satisfies
\be%
\label{apprTrans} \bar y {\rm e}^{\bar y}=A \ee
The solution to this equation is the Lambert function $\bar
y=W(A)$. It is plotted in Fig. \ref{fig2} as a function of the
initial field momentum for a fixed initial scale factor $p_0=p_*$.
We see that dependence of $\phi_{\rm max}-\phi_0$ upon $\pi_0$ is
close to logarithmic, i.e it is not very sensitive to initial
momentum. For instance, if $\pi_0=5 \lp$, the inflaton grows by
about two Planck masses during the effective phase. This is
insufficient for successful inflation.

Let us now take a non-zero coupling constant and, treating it as a
small parameter, consider the corrections to the maximum value of
the scalar field $\epsilon:=y-\bar y$. The equation
(\ref{apprSigma}) then yields
\be \label{sigmaCorrections}%
\epsilon = -\f{5 \sigma}{9 \kappa} \bar y^2 = -
\f{10}{3}\sigma(\bar \phi_{\rm max}-\phi_0)^2\ee
The first important consequence of this expression is that the
$\sigma$-corrections to $\phi_{\rm max}$ are small and
proportional to {\em minus} $\sigma$. In other words, in the case
of a negative coupling strength, the inflaton indeed gets higher
up the potential hill.

Before combining all the contributions to $\phi_{\rm max}$, we
note that there also takes place an effective (super-) inflation,
while the universe is expanding from  $a=a_i$ to $a=a_*$
\cite{ModFried}. This also provides a couple of e-folds, as the
scale factor $a \equiv \sqrt{p}$ increases by a factor of
$\sqrt{\f{p_*}{p_i}}\equiv \sqrt{2 j}$.

According to (\ref{Nquad}), $\phi_{\rm max}\approx 2.6 \mpl$ is
sufficient to provide 60 e-folds. Let us now give a numerical
estimate for the maximum inflaton value.  For $j=5$, taking the
initial conditions $p_i=p_*/2j=0.1 p_*, \phi_i=0.1 \mpl, \pi_i=5
\lp$, the effective growth of the field is $0.14 \mpl$, whereas
the classical contribution is $2.25 \mpl$. Adding everything up we
get $\phi_{\rm max}\approx 2.5 \mpl$. We shall discuss how generic
this result is and how it varies with a different choice of
initial conditions in the next section.

\section{Numerical results and discussion}
In this section, we study numerical solutions to the differential
equations (\ref{HamEqSc}). Again we will be mostly interested in
the maximum value of the inflaton for the allowed range of initial
conditions and the parameter $\sigma$.

Let us first discuss the initial field and the scale factor. As
was shown in the previous section, the dependence on the former is
rather trivial: the initial value $\phi_0$ acts as constant of
integration and should be merely added to the change in the
inflaton during the climbing phase.

\begin{figure}
\begin{center}
\includegraphics[width=7cm,height=7cm, angle=270]{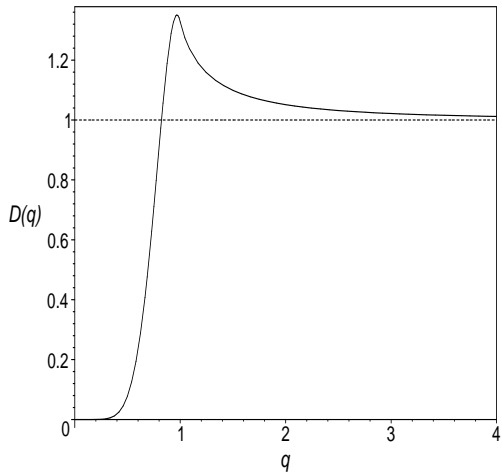}
\end{center} \vskip-0.5cm \caption{The spectrum of the geometrical
density operator $D(q)$ introduced in Eq. (\ref{defD})}
\label{fig_Dq}
\end{figure}

The dependence of $\phi_{\rm max}$ on the initial scale factor is
more complicated, but the value of $p_i$ itself is quite
restricted. It appears that there is a natural choice of $p_i$,
associated with the smallest eigenvalue of the area operator
\cite{ABL}, related to $\mu_0$. In fact, as one can see from
(\ref{phinode}), the effective growth of the scalar $\Delta
\phi_{\rm eff}$ depends on the limits of integration, hence scale
factor, only through the ratio $q\equiv \f{p}{p_*}$. The lower
limit is given by $q_i=\f{1}{2 j}$, while the upper limit is not
fixed a-priory. It has a physical meaning of the marginal value
that separates the effective and classical behavior of the
spectrum of the geometrical density operator. For the estimate in
the previous section we used $q=1$ as the upper limit. A closer
look at the graph of $D(q)$ (see Fig. \ref{fig_Dq}) shows that
there is a maximum around $q\approx 1$. In other words, the
behavior of the spectrum is not yet classical. It would be more
reasonable to stop the integration (which is meant to be over the
effective domain) at a somewhat greater value of $q$, where $D(q)$
is essentially close to unity. Recall that $q_0=1$ was giving
$\Delta \phi_{\rm eff}=0.14\mpl$. At the same time, analysis of
Eq. (\ref{phi_eff}), for $\Delta \phi_{\rm eff}$ as a function of
the upper limit of integration $q_0$, shows that the effective
growth of the inflaton sensitively depends on $q_0$, whereas the
classical formula (\ref{apprTrans}) indicates a much weaker
dependence upon $q_0$. For instance, taking $q_0=2$, one would get
$\Delta \phi_{\rm eff}=0.3\mpl$, while $q_0=4$ yields $\Delta
\phi_{\rm eff}=0.5\mpl$. This implies that if one increases $q_0$,
one should expect a somewhat greater value of total $\Delta \phi
\equiv \Delta \phi_{\rm eff} + \Delta \phi_{\rm cl}$.
Nevertheless, for the sake of estimate, we will stick with $q_0=1$
and compare the numerical results with analytical formulas.

\begin{figure}
\begin{center}
\includegraphics[width=7.5cm,height=5.0cm]{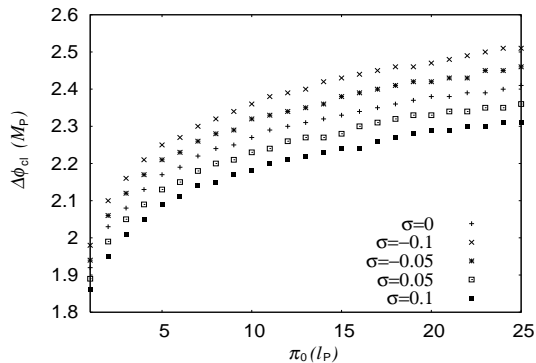}
\end{center} \vskip-0.5cm \caption{The change of the inflaton during
the classical phase as function of its initial momentum for
different values of $\sigma$ (measured in the units of $\lp^2$).}
\label{phiSc_num}
\end{figure}

Summing up, as the effective growth of the inflaton is not very
substantial and does not depend on the initial field momentum
$\pi_0$ or coupling strength $\sigma$, the quantity of the most
interest is the classical part of $\Delta \phi$. The latter is
determined by $\pi_0$ and $\sigma$.

The dependence of $\Delta \phi_{\rm cl}$ on the initial momentum
$p_0$ for five different values if $\sigma$ is displayed in Fig.
\ref{phiSc_num}. The curves correspond (top to bottom) to
$\sigma/\lp^2=-0.1,-0.05,0,0.05$ and 0.1 and are plotted for the
allowed range of initial momenta: $\f{\pi_0}{\lp}<5 j = 25$. The
maximum inflaton value grows with initial momentum and
qualitatively resembles the Lambert function. It is steepest for
small momenta and flattens out when $\pi_0$ becomes large.
Quantitatively, for $\sigma=0$ there is a very good agreement with
the analytical result (\ref{apprTrans}) and the graph of Fig.
\ref{fig2}. The curves, corresponding to opposite values of the
coupling strength, are situated symmetrically around the minimal
curve, which justifies the linear (in $\sigma$) approximation used
in (\ref{linearSigma}). Furthermore, direct calculation shows that
the $\sigma$-corrections to the central curve are very close to
those predicted by formula (\ref{sigmaCorrections}) for both small
and large initial momenta.

Classical climbing brings the scalar field fairly close to the
required values determined by $\t N = 60$ and (\ref{Nquad}). For
instance, $\phi_{\rm req}=2.6 \mpl$ for $\sigma=-0.05 \lp^2$ and
$\phi_{\rm req}=2.5 \mpl$ for $\sigma=-0.10 \lp^2$). However,
$\Delta\phi_{\rm cl}$ on its own is not enough for sufficient
inflation. At the same time, the number of e-folds is very
sensitive to deviations in $\phi_{\rm max}$ near the required
values. In other words, even small (positive) variations of
$\phi_i$ and/or $\Delta \phi_{\rm eff}$, such as of order of $0.1
\mpl$, would lead to sufficient $\t N$ and should be seriously
taken into account.

We see from Fig. \ref{phiSc_num} that a negative coupling constant
indeed gives a greater change of the inflaton, and more negative
values are preferable for successful inflation. This fact can be
also understood from the Jordanian perspective. The non-minimally
coupled action (\ref{Action}) can be thought of as a standard one
with a field-dependent gravitational constant (\ref{kappaEff}).
Clearly, a negative $\sigma$ results in `weaker' gravitational
coupling, and it is not surprising that the inflaton would reach
to a greater value (yielding a greater $N$), than for $\sigma=0$.
Moreover, if $\sigma<0$, the second term of the righthand side of
the relation between the number of e-folds in the Einstein and
Jordan frames (\ref{Nrel}) is of order one and can be neglected.
Thus $\t N \approx N$ and is greater than in the case of minimal
coupling. Similar considerations work for a positive coupling as
well and imply a smaller number of e-folds.

We should now clarify the seeming discrepancy with the result of
Futamase and Maeda \cite{Conf_NMC}. As we have already mentioned,
in that paper, the authors argued that, in order to allow
successful inflation, the coupling constant $\xi$ had to be
fine-tuned within a very narrow interval: $\xi < 10^{-12}$ for
$\xi>0$ and $|\xi|<10^{-3}$ for $\xi<0$. These restrictions on
$\xi$ arose from the heuristic argument based on Linde's chaotic
inflationary scenario. The values of $\phi_i$ are assumed to be
randomly distributed from zero up to Planckian energy densities
$V(\phi) \sim \mpl^4$. Such potential is attained at $\phi \sim
10^6 \mpl$. The latter must necessarily be less than the critical
value $\phi_{\rm crit}$ (\ref{phi_crit}), which implies $\sigma
\sim 10^{-12} \lp^2$ and $\xi \sim 10^{-14}$.

The above conditions can be relaxed in the framework of LQC. As
the initial values of the scalar field appear as vacuum
fluctuations, that are amplified during the `climbing' phase, one
just needs to restrict the coupling constant so that $\phi_{\rm
crit}$ merely exceeds $\phi \sim 3 \mpl$ - the maximum inflaton
value we are interested in. That yields $\sigma \sim 10^{-1}
\lp^2$ and $\xi \sim 10^{-3}$. The former is exactly the maximum
value of the parameter $\sigma$ we have considered in the paper.

To summarize, the main characteristic of inflation, the number of
e-folds, depends on the initial matter fluctuations and the
coupling parameter $\sigma$. The coupling strength is bounded to
be less than $0.1 \lp^2$ by the condition $\phi_{\rm
max}<\phi_{\rm crit} \equiv 1/\sqrt{|\sigma|}$ for both positive
and negative couplings. At the same time, the most negative
$\sigma$ would work best for successful inflation. The restriction
on the initial field momentum $\pi_i$ appears as a requirement for
the matter density to remain subcritical and implies $|\pi_i|<5
j$. With these restrictions, the value of the inflaton will
increase by approximately $2.0-2.6 \mpl$ during the `climbing'
(effective and classical) phase. Together with the initial vacuum
fluctuations of the scalar field of order of several tenths of
$\mpl$ this would lead to $\t N \geq 60$, i.e. sufficient
inflation. It should be noted that one cannot get $\t N$ much
greater than 60, which indicates that observations may be
sensitive to the mechanism discussed here.

{\bf Acknowledgements:} We thank P.Singh and K.Vandersloot for
valuable discussions.


\begin{thebibliography}{10}
\bibitem{Guth}
A. H. Guth, Phys. \ Rev. \ D {\bf 23}, 347 (1981).
\bibitem{L&L}
A. Liddle and D. Lyth, {\em Cosmological Inflation and Large-Scale
Structure}, Cambridge, UK: Univ. Pr. (2000).
\bibitem{Linde}
A. D. Linde, Phys. Lett. {\bf 129B}, 177 (1983).
\bibitem{Conf_NMC}
T. Futamase and K. Maeda, Phys. Rev. {\bf D39}, 399 (1989).
\bibitem{Conf_NMC1}
T.\ Futamase, T.\ Rothman, and R.\ Matzner,
\newblock Phys.\ Rev.\, {\bf D39}, 405 (1989)
\bibitem{LivRev}
Bojowald M 2005
{\em Living Rev.\ Relativity} {\bf 8} 11
\bibitem{ModFried}
M.~Bojowald, Phys. Rev. Lett. {\bf 89}, 261301 (2002).
\bibitem{2ndCorr}
K.~Vandersloot, Phys. Rev. D {\bf 71}, 103506 (2005).
\bibitem{closedinflation}
M. Bojowald and K. Vandersloot, Phys.\ Rev.\ D {\bf 67}, 124023
(2003);
\bibitem{ChaInf}
J.E.~Lidsey, D.J.~Mulryne, N.J.~Nunes, R.~Tavakol, Phys. Rev. D
{\bf 70}, 063521 (2004); G.~Date, G.M.~Hossain, Phys. Rev. Lett.
{\bf 94}, 011301 (2005); P.~Singh, Class.Quant.Grav. {\bf 22} 4203
(2005); N.J.~Nunes, Phys.Rev. D {\bf 72}, 103510 (2005); M. Kagan,
Phys. Rev. D {\bf 72}, 104004 (2005).

\bibitem{Carroll}
S.~Carroll, {\em Spacetime and Geometry}
\bibitem{nonMin}
M.~Bojowald and M.~Kagan, gr-qc/0604105
\bibitem{AshVarReell}
Barbero~G JF 1995
{\em Phys.\ Rev.\ D} {\bf 51} 5507--5510
\bibitem{Immirzi}
Immirzi G 1997
{\em  Class.\ Quantum Grav.} {\bf 14} L177--L181
%
\bibitem{geomDens}
M.~Bojowald, Phys.Rev. D {\bf 64}, 084018  (2001)
\bibitem{FieldEst}
D.J.~Mulryne, N.J.~Nunes, R.~Tavakol, J.E.~Lidsey, Int. J. Mod.
Phys. {\bf A20}, 2347 (2005)
\bibitem{dQ}
M.~Bojowald, Class.Quant.Grav. {\bf 19}, 5113 (2002)
\bibitem{SemiSV}
P. Singh and K. Vandersloot, Phys.Rev. D {\bf 72}  084004 (2005)
%
\bibitem{initFluc}
M.~Bojowald, J.E.~Lidsey, D.J.~Mulryne, P.~Singh, and R.~Tavakol,
Phys.Rev. D {\bf 70} 043530 (2004)
%
\bibitem{Date}
G.~G.Date and G.M.~Hossain, Class.Quant.Grav. {\bf 21} 4941
(2004); K.~Banerjee and G.~G.Date, Class.Quant.Grav. {\bf 22} 2017
(2005)
\bibitem{Kevin}
K.~Vandersloot, private communications
\bibitem{ABL}
A. Ashtekar, M. Bojowald, and J. Lewandowski, Adv. Theor. Math.
Phys. {\bf 7}, 233 (2003)

\end{thebibliography}
\end{document}